# High-resolution digital control of highly multimode laser

Chene Tradonsky, Simon Mahler,* Gaodi Cai, Vishwa Pal, Ronen Chriki, Asher A. Friesem, and Nir Davidson
*Department of Physics of Complex Systems, Weizmann Institute of Science, Rehovot 7610001, Israel*

**Abstract:** A rapid and efficient method for generating laser beams with controlled intensity, phase and coherence distributions is presented. It is based on a degenerate cavity laser in which a digital phase-only spatial light modulator is incorporated. We show that a variety of unique and high-resolution shaped laser beams can be generated with either a low or a high spatial coherence. We also show that by controlling the phase, intensity and coherence distributions, a laser beam can be efficiently reshaped after propagation.

**1. Introduction**

In general, it is difficult to control the intensity, the phase and the coherence distributions of a light source simultaneously [1, 2]. Such control would allow the development of laser outputs with arbitrary shaped distributions and properties that are needed for many applications [2-11]. So, many efforts have been extended toward achieving this control. These included separate control of the laser's intensity and phase and separate control of the laser's coherence.

The separate control of the spatial coherence was done rapidly and efficiently by means of a degenerate cavity laser (DCL) [12] and an intra-cavity spatial filter aperture [13], demonstrating that the number of independent lasing modes can range from 1 to 320,000 with less than a 50% change in output power [14]. It was exploited for wide field speckle-free illumination and imaging [15]. Recently, by adding an intra-cavity phase only diffuser into the DCL, the spatial coherence of the laser was rapidly (ultrashort nanosecond time regime) reduced [16].

The separate control of the intensity and phase distributions was investigated with several laser beam-shaping techniques [1-5, 17]. An efficient and flexible method involves the use of intra-cavity dynamic diffractive element implemented by means of a spatial light modulator (SLM) [18]. Due to mode competition and dissipative coupling [19, 20], the intensity and phase distributions were efficiently controlled with low loss [18]. However, due to the low number of lasing modes in the laser cavity, the method was limited to patterns with relatively low resolution.

By combining a DCL, an intra-cavity SLM and an intra-cavity spatial Fourier aperture, it is possible to exploit a very large number of independent lasing modes of the DLC and have direct access to both the near-field and far-field planes for independent manipulations and control of several degrees of freedom of the lasing beam. Such approach was used for rapidly solving the phase retrieval problem [10] and is reminiscent to that used for remote structuring in waveguides [8].

Here we present a rapid and efficient method to generate laser beams with almost arbitrary intensity, phase and coherence distributions. It is based on a digital degenerate cavity laser (digital DCL) in which a phase only spatial light modulator (SLM) is incorporated. With such method, we generated a variety of unique and high-resolution shaped laser beams with both high and low spatial coherence. We investigated the tradeoff between coherence and spatial resolution, and showed that a laser beam with coherent light has a lower resolution than one with incoherent light, but the one with coherent light is more propagation invariant. We also generated lasing at pure Hermite-Gaussian modes of extremely high orders (up to $HG_{50,50}$) with good efficiency and accuracy. Finally, we show how to shape the beam at the lasing output such that it is transformed into an arbitrary desired and substantially different shape by free-space propagation only.

*Email: sim.mahler@gmail.com

## 2. Experimental arrangement

To demonstrate and study the digital degenerate cavity laser, we resorted to two experimental arrangements, a simple linear digital DCL and a more elaborated ring digital DCL, both based on the same operation principle and both can be used for most of our experiments [21]. The ring digital DCL, which allows greater versatility because the light can propagate unidirectional inside the cavity, is more complex and is described in some detail in the Supplementary Material.

Figure 1 schematically shows the linear digital DCL arrangement. At one end, a reflective phase only spatial light modulator (SLM) serves as a back mirror and at the other end a ND:YAG gain medium is adjacent to an output coupler. A $4f$ telescope, formed of lenses Lens1 and Lens2 of focal length $f$, is midway in the cavity a distance $f$ away from the SLM and the output coupler. The $4f$ telescope precisely images the field distribution at the output coupler plane onto the SLM plane and vice versa, which are denoted as the near-field planes and are equivalent. The plane between Lens1 and Lens2 is denoted far-field plane, where a Fourier aperture is placed. A Fourier relation exist between the near-field and far-field planes. An imaging system (not shown) after the output coupler images and detects both the near-field (SLM) and the far-field (intra-cavity aperture) planes onto a camera. More details are in the Supplementary Material.

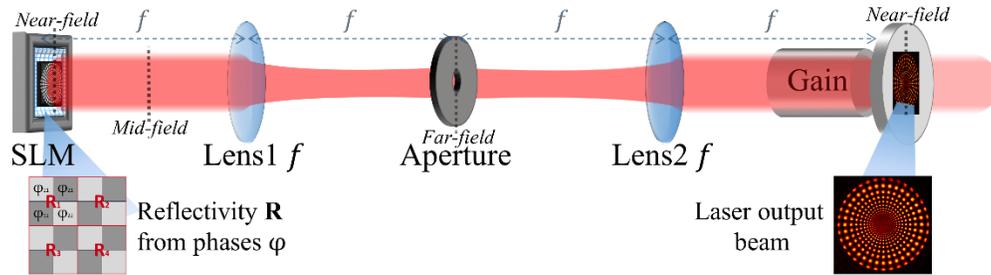

**Fig. 1.** Digital linear degenerate cavity laser arrangement. A laser output beam with a desired distribution is obtained by controlling the intensity and phase distributions inside the laser cavity by means of the digital spatial light modulator (SLM) at the near-field plane and an adjustable intra-cavity aperture at the far-field plane. The distribution of the lasing modes and their number can thus be controlled. An additional intra-cavity aperture located at a mid-field plane (midway between the near-field and the far-field) can be used to control simultaneously the phase, intensity and coherence distributions (see Fig. 5).

The SLM at the near-field serves as a mirror. Its reflectivity and phase distributions are both locally controlled by manipulating the phase difference between $2 \times 2$ adjacent pixels so as to form a super-pixel scattering light (inset at SLM) and then filtering the relevant light with the far-field aperture [13, 18, 22, 23]. The control of the reflectivity distribution of the SLM allows control of the intensity distribution of the laser beam in the near-field, whereas control of the phase distribution of the SLM allows control of the frequency distribution. Each super-pixel acts as an independent mirror and all the pixels together form an array of incoherent and independent lasing super-modes [21]. By varying the size and shape of the far-field aperture, it is possible to control the phase distribution and coherence of the laser light [14, 15].

## 3. Experimental results

Using the digital DCL arrangements, we performed a series of experiments to demonstrate control of the intensity, phase and spatial coherence distributions of their output light. Representative results are presented in Figs. 2–5.

Figure 2 shows two controlled intensity distributions at the output of the digital DCL operated high above its lasing threshold. As evident, the intensity distribution of the laser output beam is nearly identical to the gray scale high-resolution images displayed on the SLM (inset). This is because, high above lasing threshold, the lasing power of each super-pixel is proportional to its reflectivity. The high resolution is provided by the very high number of independent lasing modes in the DCL, which can reach 320,000 [14]. With such number of independent lasing modes, the resolution is in principle limited by the resolution of SLM (20 $\mu$m/pixel) where each lasing super-pixel in the SLM corresponds to a combination of modes (super-mode). Accordingly, the intensity of each super-mode was controlled by the reflectivity of the SLM while the phase was randomly distributed. These results indicate that it is possible to control the intensity distribution of a lasing beam with high resolution, and correspondingly the mode distribution of the DCL. Due to the high number of independent lasing modes and their random phase distribution, the laser light was fully incoherent. When the DCL operates close its lasing threshold (where the relation between reflectivity and lasing power becomes nonlinear), or when the different super-pixels are significantly coupled (e.g. with a far-field aperture), or to correct for inhomogeneous gain profile, any desired lasing intensity distribution can still be obtained by resorting to an iterative correcting procedure of each super-pixel reflectivity, see Supplementary Material.

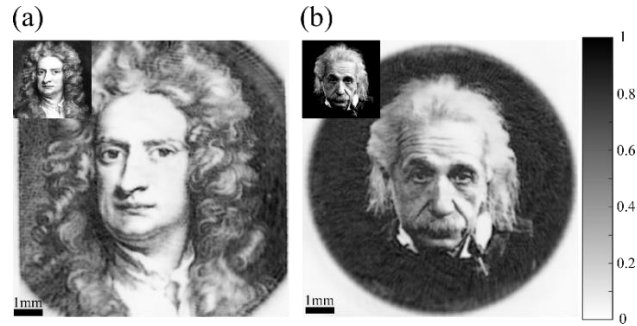

**Fig. 2.** Experimental intensity distributions from the digital degenerate cavity laser. (a) Incoherent Newton shaped beam –"Newton mode"; (b) Incoherent Einstein shaped beam – "Einstein mode". Insets – images displayed on the spatial light modulator.

Next, we investigated how the free space propagation of the laser output beam is affected by the coherence of the light. The results, presented in Fig. 3, show the intensity distributions of the laser beam at the near-field plane ($z = 0$ mm) and after propagating the beam to a distance $z = 12.5$ mm, for incoherent light and partially coherent light. Figure 3 (a) shows the results for fully incoherent light (as used in Fig. 2). The intensity distribution at $z = 0$ mm is shaped according to a sector resolution target, composed of circular distributions of spots, whose size becomes smaller closer to the center. After free space propagating $z = 12.5$ mm, the intensity distribution is significantly distorted where each spot broadens, as expected for fully incoherent light. Figure 3 (b) shows the results for laser light with higher coherence, obtained by inserting a small pinhole aperture in the far-field (Fourier) plane of the DCL [13, 14]. As evident, the intensity distribution at $z = 0$ mm is similar to that in Fig. 3(a), albeit with a somewhat lower resolution. After propagating $z = 12.5$ mm, the intensity distribution remains almost the same.

The results of Fig. 3 reveal a tradeoff between spatial resolution and free-space propagation without distortion. When the laser light is fully incoherent, Fig. 3(a), the intensity distribution at $z = 0$ mm has higher resolution, than when the laser light is partially coherent, Fig. 3(b). but also distorts much faster upon free-space propagation. This is because the small far-field aperture suppresses the high spatial frequency components of the laser coherent light, thus reducing the resolution of the intensity distribution at $z = 0$ mm and also reducing the distortion due to free-space propagation. Previously, we have shown that if the intra-cavity far-field circular aperture is replace with a thin annual aperture, the distortion of an arbitrary image due to free space propagation can be further reduced [24].

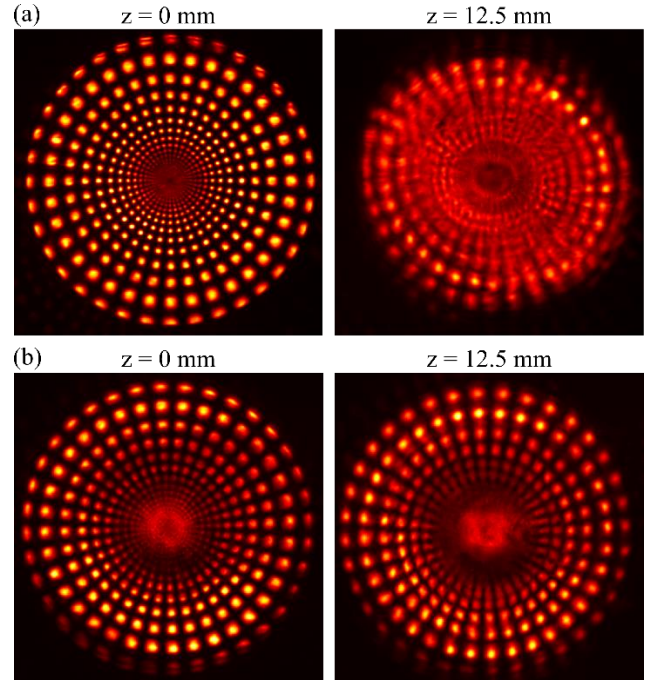

**Fig. 3.** The effect of coherence on the intensity distribution of a propagating laser beam generated with a digital degenerate cavity laser. (a) Intensity distribution with an incoherent laser light beam at $z = 0$ mm and $z = 12.5$ mm (no intra-cavity aperture). (b) Intensity distribution of a more coherent laser light beam at $z = 0$ mm and $z = 12.5$ mm (4 $mm$ diameter aperture in the intra-cavity far field). The coherence of the laser light was continuously controlled by varying the diameter of the intra-cavity far-field aperture (not shown).

Next, we investigated the generation of very high order Hermite-Gaussian modes [25]. They generally have a square shape with sharp edges and retain the same shape after Fourier transformation [25, 26]. First, we generate the intensity distribution of any desired Hermite-Gaussian $HG_{n,m}$ beam of order n and m using an intra-cavity SLM [18]. To impose the correct phase distribution, a square aperture that tightly matches the beam size of the desired Hermite-Gaussian $HG_{n,m}$ was inserted in the far-field plane of the digital DCL. This matched far-field aperture ensured that the correct phase distribution corresponds to the minimal loss mode, whereby the output lasing mode from the digital DCL is the desired Hermite-Gaussian due to mode competition, as in the phase retrieval approach [21].

The results for Hermite-Gaussian beam of order $HG_{10,10}$ (with 121 near-Gaussian spots) are presented in Fig. 4. Figure 4 (a) and (c) show the measured near-field and far-field intensity distributions of the lasing output beam. The phase distributions in the near-field and far-field (Fig. 4(b) and (d)) were reconstructed from the measured intensity distributions by using a Gerchberg–Saxton phase retrieval algorithm [27]. As evident, both intensity distributions agree well with a Hermite-Gaussian beam of order $HG_{10,10}$ and both phase distributions show the expected $\pi$ phase difference between neighboring spots manifesting single and pure lasing mode of very high order. Hermite-Gaussian lasing modes of even higher orders, up to $HG_{50,50}$ (with 2601 near-Gaussian spots), were also generated, albeit with a somewhat lower purity (Supplementary Material).

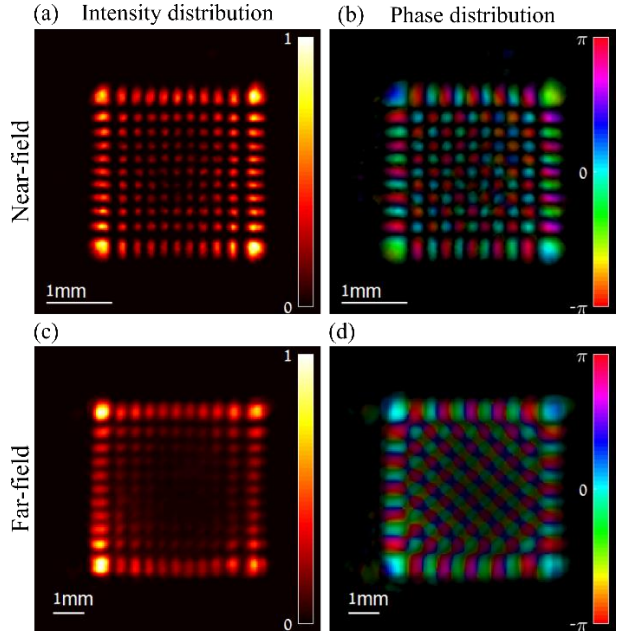

**Fig. 4.** Experimentally detected Hermite-Gaussian beam $HG_{10,10}$ generated with the experimental arrangement of the digital DCL. Near-field (a) intensity and (b) phase distributions of the generated $HG_{10,10}$ laser beam. Corresponding far-field (c) intensity and (d) phase distributions of the generated $HG_{10,10}$ laser beam.

Finally, we demonstrate how to obtain a lasing beam with two different intensity distributions at two different propagation distances. First, we imposed one desired intensity distribution at the near-field plane ($z = 0$ mm) by controlling the reflectivity of the intra-cavity SLM, as discussed above. Next, we inserted an additional shaped aperture in the digital ring DCL at a mid-field plane, located at a distance $z_{MD}$ away from the near-field plane.

Due to the Fresnel diffraction, the diffracted pattern at a distance $z_{MD}$ away from near-field depends both on the intensity and on the phase distributions of the near-field. While the near-field intensity distribution is constrained by the reflectivity of the SLM, the laser is free to choose the phase distribution that minimizes loss imposed by the mid-field aperture, i.e. to generate at the mid-field plane a shape consistent with the mid-field aperture. The mid-field aperture could be replaced with a transmissive SLM in order to obtain variable control.

We demonstrated this scheme by imposing a "superman" image at $z = 0$ mm by the near-field SLM and imposing a "batman" image using a "batman" shaped aperture located at a distance $z_{MD}$. The results, showing the intensity distributions at four different propagation distances $z = 0$ mm (near-field plane), $z = 100$ mm, $z = 200$ mm and $z = z_{MD} = 300$ mm (mid-field plane) are presented in Fig. 5. As evident, the distributions gradually change from the good quality "superman" image at the near-field plane (Fig. 5(a)), to poor quality at intermediate planes [Figs. 5(b) and 5(c)], to good quality "batman" image at the mid-field plane (Fig. 5(d)).

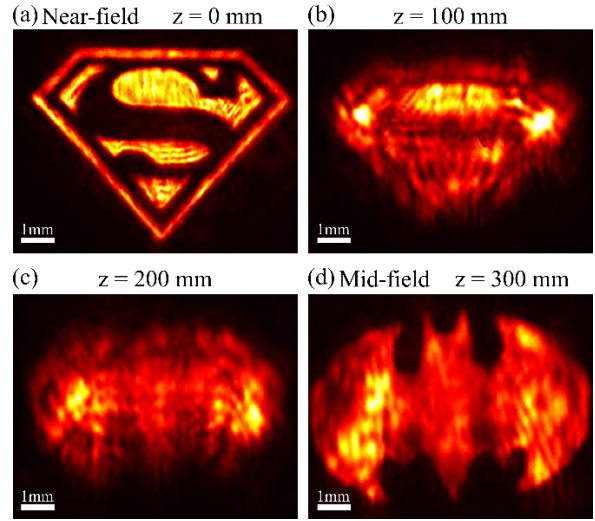

**Fig. 5.** Intensity distributions at different propagation distances. (a) At near-field plane $z = 0$ mm, (b) $z = 100$ mm, (c) $z = 200$ mm and (d) mid-field plane $z = 300$ mm.

### 4. Concluding remarks

We presented a rapid and efficient method for generating laser beams with controlled intensity, phase and coherence distributions. It involved a digital degenerate cavity laser in which a phase only spatial light modulator and spatial filters are incorporated. A variety of unique and high-resolution shaped laser beams were demonstrated with both low and high spatial coherence, as well as pure Hermite-Gaussian beams of very high orders. We observed a tradeoff between coherence and spatial resolution, whereby the image of a shaped laser beam with coherent light had lower spatial resolution than one with incoherent light, but was much less distorted by free-space propagation. We also demonstrated that by controlling the phase, intensity and coherence distributions simultaneously, a shaped laser beam at one plane can be efficiently reshaped after free-space propagation into a completely different shape.

We believe that our method and results sufficiently indicate that it is now possible to generate arbitrary laser output distributions, which could lead to new and interesting applications. Such shaped laser beam can be generated within less than 1 µs [21] and with less than 15% reduction in output power [16].

**Acknowledgement.** The authors thank the Israel Science Foundation (ISF) (Grant No. 1881/17) for its support.

# Supplementary Material:

## I. Methods

In this section, we first introduce a basic digital linear degenerate cavity laser (Fig. S1), then a simple digital ring degenerate cavity laser (Fig. S2) and finally an actual detailed digital ring degenerate cavity laser that was used in our experiments (Fig. S3). We also discuss the pros and cons of each laser configuration and present the technique for using a phase only spatial light modulator as intensity and phase mirror.

### (a) Digital linear degenerate cavity laser

Figure S1 schematically shows the arrangement of a digital linear degenerate cavity laser. It is comprised of a spatial light modulator (SLM) that serves as a back mirror at near-field plane, a gain media, two lenses in a 4f telescope configuration and a front mirror. Lens1 and Lens2 form a $4f$ telescope inside the cavity, which precisely images the field distribution at the back mirror (SLM) plane onto the front mirror plane and vice versa. Those planes are denoted near-field planes and are equivalent. The plane midway between Lens1 and Lens2 (focal plane) is denoted far-field plane. A Fourier relation exist between the near-field and far-field planes, and both planes are physically accessible. Such a degenerate cavity laser has a huge number of lasing modes, whose the number can be varied by controlling the size of an aperture at the far-field plane, and thereby the spatial coherence of the laser.

The SLM serves as a mirror where both its reflectivity and phase distributions are controlled locally. The control is done by manipulating the phase difference between $2 \times 2$ adjacent pixels so as to form a super-pixel, shown in the left inset of Fig. S1, from which light is scattered. The scattered light is then filtered by an aperture at the far-field Fourier plane. The control of the reflectivity distribution thereby allows control of the intensity distribution of the laser beam in the near-field, as shown in the example in the right inset of Fig. S1, whereas control of the phase distribution allows control of the frequency distribution. Each super-pixel serves as an independent mirror and all the pixels together form an array of incoherent and independent lasing super-modes. Some spatial light modulators only operate at a specific polarization of laser light, so a wave plate or a polarizer must be inserted in the degenerate cavity laser in front of the spatial light modulator.

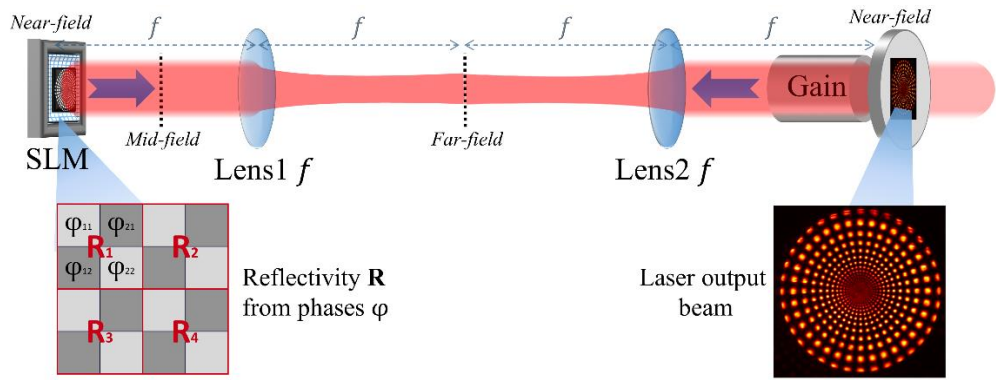

Figure S1: Arrangement of a digital linear degenerate cavity laser. The SLM denotes spatial light modulator, blue arrows the direction of light propagation, and $f$ the Focal length of the lenses.

In order to avoid extraneous scattering and thereby extra loss, the gain media in Fig. S1 was placed next to the front mirror (front near-field). Alternatively, the gain media can be placed next to the spatial light modulator at the back near-field and produce similar results (not shown). Lens1 and Lens2 can also have two different focal lengths. If so, there will be a magnification ratio between the front and back near-field planes and the far-field plane will be at the focal plane between the two lenses.

The linear degenerate cavity laser has the advantage that it is relatively easy to align. However, the light must propagate forth and back inside the laser cavity every round-trip, so it passes through each element twice. Due to the $4f$ telescope and mirror configuration, the laser field on the way-back is rotated by 180°. Thus, careful rotation alignment and symmetry considerations are needed. These can be avoided by resorting to ring degenerate cavity lasers.

(b) Simple digital ring degenerate cavity laser

A digital ring degenerate cavity laser is similar to a digital linear degenerate cavity laser except that the light propagates in only one direction and in a single pass inside the laser cavity. For that, a second $4f$ telescope is inserted and four mirrors oriented at 45° connect them. A typical arrangement for a simple digital ring degenerate cavity laser is schematically shown in Fig. S2. In order to implement unidirectional propagation and a single pass, one laser light propagating direction must be blocked, e.g. with an optical isolator (not shown). Note also that a unidirectional ring laser suppresses spatial hole burning.

The arrangement in Fig. S2 has some problems. First, the spatial light modulator (SLM) must be transmissive. For such an SLM the zero-order efficiency is typically low and the response time relatively slow. Second, due to the four mirrors at 45°, a single distance between two optical elements cannot be changed separately. For these reasons, we resorted to a more practical experimental arrangement, described in the next section.

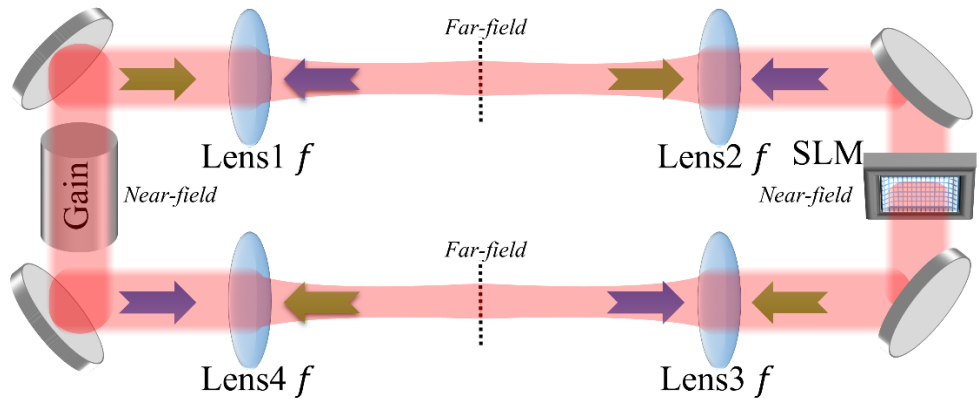

Figure S2: Experimental arrangement of a digital ring degenerate cavity laser The SLM denotes spatial light modulator, blue or green arrows directions of light propagation.

(c) Actual digital ring degenerate cavity laser

The arrangement for an actual and practical digital ring degenerate cavity laser that we used in our experiments is schematically shown in Fig. S3. It is comprised of two $4f$ telescopes formed of one common lens of focal length $f_2$ and two separate lenses of focal lengths $f_1$, a reflective phase only spatial light modulator (SLM), a Faraday rotator and two polarizing beam splitters (PBS), two half wave plates at 45° and at 22.5°, two retroreflectors, a pentaprism-like 90° reflector and a gain medium with anti-reflective coating on both ends. In addition, an aperture of arbitrary shape and size, can be can be inserted in the far-field plane.

The gain medium at one end of the cavity is a 1.1% doped Nd:YAG rod of 1 cm diameter and 11 cm long , lasing at $\lambda = 1064$ nm. It is optically pumped by a quasi-CW 100 μs duration Xenon flash lamp operating at 1 Hz to minimize thermal lensing effects. The gain medium is placed at focal distance $f_1 = 50$ cm away from a Lens1 of 5.08 cm diameter. A second Lens2 of focal length $f_2 = 75$ cm and 5.08 cm diameter is inserted at distance $f_1 + f_2$ away from Lens1. Between Lens2 and Lens1, a half-wave plate at 45° horizontally polarizes the light such that all the light passes through a polarized beam splitter (PBS1) with no orthogonal reflection.

After Lens2, a reflective phase only spatial light modulator was inserted at focal distance $f_2 = 75$ cm away from Lens2 and acted as a mirror. Between the spatial light modulator and Lens2, a Faraday rotator together with a half-wave plate and a polarized beam splitter (PBS2) acted as an optical isolator and ensured unidirectionality of the laser light inside the cavity. The laser light is reflected back by the SLM and propagates back through PBS2, the half-wave plate, the Faraday rotator, Lens2 and PBS1. During the passage back, a small percentage (5%) of the light reflected by PBS2, acting as output coupler. Due to the Faraday rotator and half-wave plate ensemble, the light was vertically polarized and was thereby orthogonally reflected by the polarized beam splitter PBS1. A Lens3 of 5.08 cm diameter and focal length $f_1$ is inserted at a distance $f_2 + f_1$ from Lens2 and at a distance of $2f_1$ from Lens1. Midway between Lens2 and Lens3 (Fourier far-field plane), a varying size aperture is be inserted. All the lenses were plano-convex spherical lenses with anti-reflection coating for 1064 nm. The spatial light modulator was a Hamamatsu LCOS-SLM X13138-03 with 98% reflectivity at 1064 nm and high damage threshold.

Finally, an external imaging system, shown in Fig. S3b, is used to simultaneously detect both the near-field and the far-field Fourier planes onto a camera.

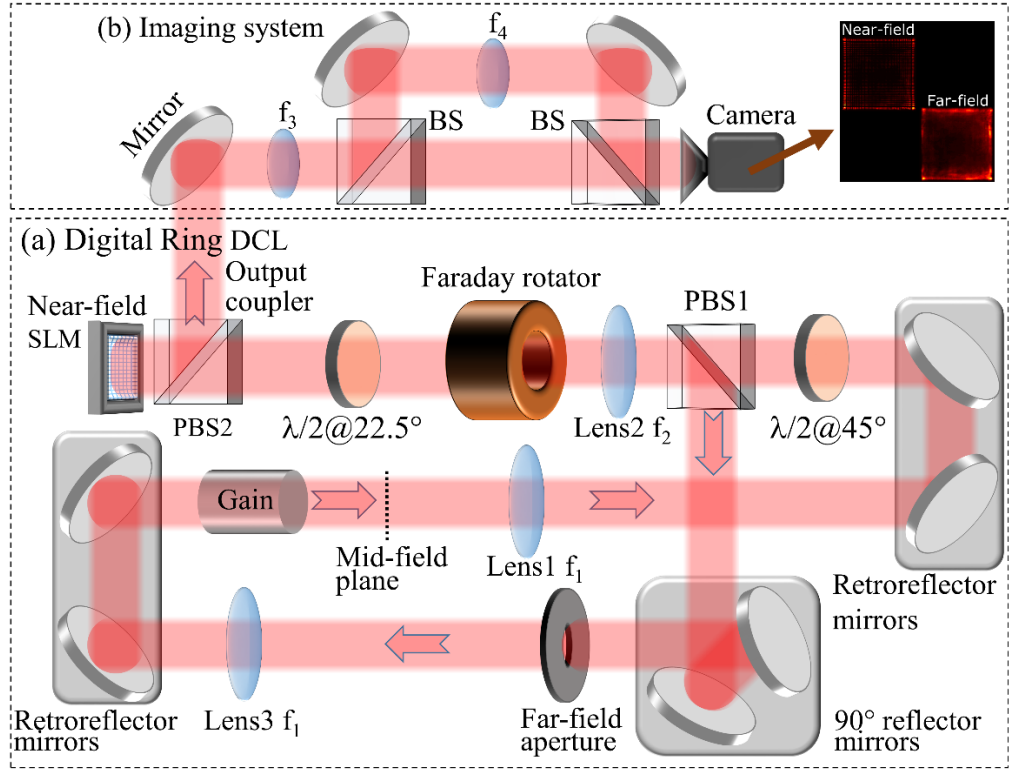

Figure S3: Experimental arrangement used to perform the experiments. (a) Digital ring degenerate cavity laser. A Faraday rotator together with two polarizing beam splitters PBS1 ensure unidirectional light propagation, which together with ring arrangement allow the use of asymmetrical elements without symmetry consideration. (b) Imaging system. SLM: Spatial light modulator.

## II.     Hermite-Gaussian $HG_{20,20}$, $HG_{30,30}$ and $HG_{50,50}$ laser beams

Figure 4 of the main text showed the near-field and far-field intensity and phase distributions of a Hermite-Gaussian $HG_{10,10}$ laser beam with 121 spots. In addition, we also generated Hermite-Gaussian $HG_{20,20}$, $HG_{30,30}$, and $HG_{50,50}$ laser beams as shown in Fig. S4. As evident they are distinguishable as Hermite-Gaussian beams of high orders.

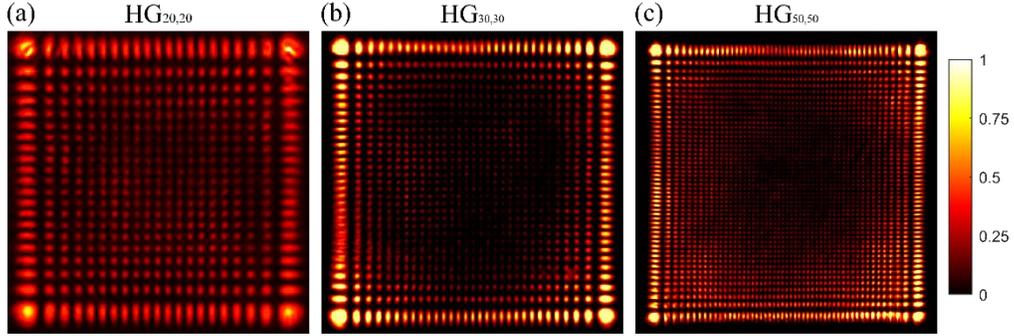

Fig. S4. Experimentally detected near-field intensity distributions of Hermite-Gaussian laser beams generated with the digital degenerate cavity laser. Hermite-Gaussians (a) $HG_{20,20}$, (b) $HG_{30,30}$, (c) $HG_{50,50}$.

## III.  Correction of non-uniformities in the lasing intensities

As shown in Fig. 5 of text, the quality of some intensity distributions is poor. This is due to laser intensity non-uniformities which could be corrected by a feedback algorithm, implemented on the spatial light modulator, as we explain below. Since we only had one SLM, only one intensity distribution was corrected.

There are several sources of laser intensity non-uniformities. These include the non-uniformities of gain media, spherical aberrations of the lenses, noise coming from the flash lamp, and phase distortions of the spatial light modulator. To correct the non-uniformities, we implemented a feedback algorithm to correct the reflectivity distribution of the near-field spatial light modulator.

The intensity feedback applied on the reflectivity $R_i^{(n)}$ at the $i$th pixel of the spatial light modulator after $n$ iterations is $R_i^{(n)} = R_i^{(n-1)} + p \left( \left| E_{\text{ref}}(\vec{k_i}) \right|^2 - I_{\text{meas},i}^{(n-1)} \right)$ where $p$ is a feedback parameter, $I_{\text{meas},i}^{(n-1)}$ is the measured intensity at the $i$th SLM pixel after $n-1$ iterations, and $\left| E_{\text{ref}}(\vec{k_i}) \right|^2$ is the input reference intensity at the $i$th pixel.

Figure 8 shows the improvement of the laser intensity after applying the spatial light modulator feedback algorithm. Figure S5 (a) shows the near-field intensity distribution of a shaped laser beam in the digital degenerate cavity laser with a near-field spatial light modulator and no far-field aperture. Due to the high number of independent lasing modes and low spatial coherence, the intensity distribution of the beam is correctly controlled with a high resolution. Figure S5 (b) shows the near-field intensity distribution after inserting a relatively small far-field aperture. As evident, the intensity distribution is less uniform due to the lower number of independent modes and the increase of the spatial coherence. Figure S5 (c) shows the near-field intensity distribution after using the spatial light modulator feedback algorithm. As evident, the intensity distribution is more uniform, improving the quality of the image.

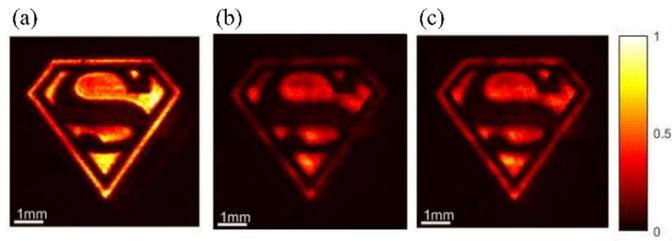

Figure S5. Optimization of a laser beam intensity distribution using a feedback algorithm on the spatial light modulator in the digital degenerate cavity laser. (a) Near-field intensity distribution of the desired beam without a far-field aperture. (b) Near-field intensity distribution after the insertion of a 4 mm diameter aperture in the far-field. (c) Near-field intensity distribution after using the spatial light modulator feedback algorithm.